\documentclass[5p]{elsarticle}
\pdfoutput=1
\newcommand{\email}{\ead}
\newcommand{\affiliation}{\address}
\newcommand{\onlinecite}{\cite}

\usepackage{graphicx}
\usepackage{amssymb}
\usepackage{amsmath}
\usepackage{enumitem}
\usepackage{textcomp}
\usepackage{gensymb}
\usepackage{mciteplus}

\newcommand{\ud}{\,{\mathrm{d}}}
\newcommand{\zc}{\overline{z}}
\newcommand{\wc}{\overline{w}}

\newcommand{\uRe}{{\mathrm{Re}}\,}

\newcommand{\uMs}{M_{\mathrm{S}}}

\DeclareMathOperator{\sech}{sech}

\def\be{\begin{equation}}
\def\ee{\end{equation}}

\begin{document}

\title{What makes magnetic skyrmions different from magnetic bubbles ?}

\author{Andrei B. Bogatyr\"ev}
\affiliation{Institute for Numerical Mathematics, Russian Academy of Sciences, 8
 Gubkina str., Moscow GSP-1, Russia 119991}
\email{ab.bogatyrev@gmail.com}

\author{Konstantin L. Metlov}
\affiliation{Donetsk Institute of Physics and Technology, 72 R. Luxembourg str.,
 Donetsk, Ukraine 83114}
\email{metlov@fti.dn.ua}
\date{\today}

\begin{abstract}
A large enough piece of ferromagnet is usually not magnetized uniformly, but develops a magnetization texture. In thin films these textures can be doubly-periodic. Such are the well known magnetic bubble domains and the recently observed ``skyrmion'' magnetization textures in MnSi. In this paper we develop a theory of periodic magnetization textures, based on complex calculus to answer the question -- is there a difference between those two textures even if they seem to carry the same topological winding number (or topological charge) ? We find that such difference exists, facilitated by a different role played by the magnetization vector's in-plane phase. We separate classical-like and quantum-like features of magnetization textures and highlight the role of magnetic anisotropy in favouring either of these cases.
\end{abstract}

\maketitle

There is a renewed interest to the topological properties of magnetization textures, which goes in concert with the recent developments in the theory of quantum Hall effect and the discovery of topological insulators. The idea of such topological states can be tracked back to the pioneering works by A. Skyrme \cite{S58,Skyrme61,Skyrme62} who have proposed to use the non-linear $\sigma$-model to describe hadrons and found its singular point-like solution called ``hedgehog'' or later ``skyrmion''. Next, the multi-skyrmion solutions of this model in two dimensions were found \cite{BP75} by identifying them mathematically with mappings of a sphere onto itself. They were called ``Belavin-Polyakov solitons'' for a time and later became a part of a larger family of ``topological solitons'' highlighting the importance of topology in stabilizing these configurations and encompassing objects of any dimension, such as one-dimensional ``domain walls'' or ``kinks''. Behind them is the central concept of the topological winding number (or ``topological charge'') -- an integer quantity, which is conserved in the non-linear $\sigma$-model. It can be explained as a number of times the Riemann sphere, representing the magnet's plane with a single ``infinitely distant'' point, covers the sphere, corresponding to the endpoint of a fixed-length (magnetization) vector. The topological charge conservation of $\sigma$-model makes it tempting to conclude that just by possessing the topological charge a certain configuration of fields will inherently be absolutely stable (and impossible to destroy). In Nature, however, the topologically non-trivial fields are created and destroyed all the time. In theory too, the significance of various properties of the field configuration (such as topological charge) exists only in the context of the model, governing the field evolution and equilibrium. There are field configurations with the topological charge, which {\em is not} conserved. Strictly speaking, they are not topological solitons or skyrmions, even if they have particle-like aspects to their evolution. Magnetic domains (and magnetic bubbles \cite{bobeck1975bubbles} in particular) are well known examples of such objects, which can easily be created and annihilated at any point of a magnetic film and not necessarily in pairs \cite{eschenfelder1980}. This leads us to the main question of this work -- can we pinpoint a fundamental difference between the skyrmions and magnetic bubbles ?

In thin ferromagnetic films, where the skyrmions and bubbles exist, the four magnetization field configurations --  an isolated 2-d hedgehog, magnetic vortex\footnote{In the literature the term ``magnetic vortex'' is also often used to denote topological solitons in magnetic nanostructures. There should be no confusion since the infinite thin films considered here do not support meron spin configurations.} (if all spins in the hedgehog are rotated by 90\degree in the film plane) and a magnetic bubble with either N{\'e}el or Bloch domain wall are all topologically equivalent. However, we know of a wealth of phenomena, such as \cite{MalozemoffSlonczewski} transitions between different domain wall types (all topologically equivalent), which are impossible if we assume them to be the same, based on topological equivalence alone. Thus, a difference we're seeking not only exists, but also can bring fruits in terms of new physical phenomena.

In addition to the topological charge conservation, there are other properties of the magnetic skyrmions, which are due to the topology of the magnetic system (or, more specifically, follow from the magnetization vector field continuity and the boundary conditions), which we call the ``topological constraints''. We will use these properties for singling out the skyrmion magnetization textures (and physical systems, supporting them) to help focus on novel and largely unexplored phenomena involving the interplay of strong non-linearity and topology.


A good case for our study is provided by the periodic magnetization textures in thin films. The periodic magnetic bubble lattice is well known, its theory \cite{Thiele:1969:TCM} have existed since the second half of the last century. The periodic lattice of skyrmions was recently reported in MnSi \cite{yu2010realspaceobservation}. To compare these structures we first develop a theory of periodic magnetization textures under the similar assumptions to those by A.~A.~Belavin and A.~M.~Polyakov. Then we derive a set of constraints of topological nature stemming from the continuity of the textures and the specific boundary conditions (periodic in this case). Next we consider several examples (both theoretical and experimental) of periodic magnetization textures in light of these constraints and draw our conclusions at the end.

The magnetization texture $\vec{M}(\vec{r})$ in a ferromagnet is a vector field, resulting from minimization of several energy terms (in this work we deal only with quasistatic case). The most important of them is the exchange interaction, which is responsible for ferromagnetism (existence of the local spontaneous magnetization $|\vec{M}(\vec{r})|=\uMs(T)>0$ when the exchange interaction is strong enough to suppress the paramagnetic temperature fluctuations at temperature $T<T_C$) and for the topological features of the magnetization vector field $\vec{M}(\vec{r})$. The latter is true both in the field theory (as follows from the A. Skyrme's line of work) and for ferromagnets \cite{BP75}, where the exchange energy density in the simplest isotropic case is proportional to $|\vec{\nabla} M_\mathrm{X}|^2+|\vec{\nabla} M_\mathrm{Y}|^2+|\vec{\nabla} M_\mathrm{Z}|^2$ with
$\vec{\nabla}$ being the gradient operator and the indices denoting the Cartesian components of the vector $\vec{M}$. The rest of energy terms (such as crystalline magnetic anisotropy, Zeeman energy, magnetic dipolar interaction) are very important too for stabilization/destabilization of certain magnetization patterns and in general for determining the scale for the magnetization texture (since the exchange-only variational problem is scale-invariant). They will, at the very least, produce a certain distortion of the exchange-only patterns. Nevertheless, since the topological features are usually insensitive towards small distortion and even rescaling, we will be following up directly on the Belavin and Polyakov work \cite{BP75} and consider the exchange-only solutions. Thus, our analysis is bound to be qualitative with respect to specific form of the magnetization distributions, but the topological properties we are hoping to find can be expected to be valid much further beyond our simple approximations. 

Specifically, for an infinite thin film lying in the $X$-$Y$ plane of a Cartesian coordinate system, if we introduce the normalized magnetization vector $\vec{m}=\vec{M}/\uMs$, the constraint $|\vec{m}|=1$ can be automatically satisfied by the magnetization vector field components (under assumption that the film is thin enough to ignore their $Z$-dependence) expressed via stereographic projection
\begin{eqnarray}
m_X+\imath m_Y = \frac{2 w(z,\zc)}{1+w(z,\zc)\wc(z,\zc)} \label{eq:magveccomp} \\
m_Z = \frac{1-w(z,\zc)\wc(z,\zc)}{1+w(z,\zc)\wc(z,\zc)}, \nonumber
\end{eqnarray}
where $z=X+\imath Y$ is a complex coordinate, $\imath=\sqrt{-1}$, over-line denotes the complex conjugation (e.g. $\zc=X-\imath Y$) and the function $w(z,\zc)$ is an arbitrary complex function of complex variable (not necessarily analytic). The main result of Belavin and Polyakov \cite{BP75} (which is especially evident in complex notation by Gordon Woo \cite{Woo77}) after performing the variational minimization of the exchange energy is that meromorphic functions $w(z,\zc)=f(z)$ correspond to metastable states of an infinite thin film ferromagnet. Doubly periodic magnetization textures are then described by doubly periodic meromorphic functions $f(z)$.

After rescaling and rotation we can represent every 2-d lattice as generated by the lattice vectors (in complex notation) $1$ and $\tau$ with $Im~\tau>0$:
$$
L=L(\tau):=\{M+N\tau, \qquad M,N\in {\mathbb{Z}}\},
$$
where $\mathbb{Z}$ is the set of integers. There is a substantial freedom in choosing an elementary cell of the lattice, which tiles the plane when translated by all the elements of $L$. For instance, it can be the parallelogram
$$
\Pi:= \{z=t_1+t_2\tau, \quad 0<t_s<1, ~~s=1,2\}
$$
or any translation of it or -- in case of a hexagonal lattice
($\tau=\exp{\imath\pi/3}$) -- a regular hexagon.

Zeros and poles of $f(z)$ correspond to vortices and anti-vortices (or saddles) of the magnetization pattern. Chirality $\chi_j$ of a $j-th$ vortex with the center at $z=z_j$ can be measured by the first order Taylor expansion coefficient of $f(z)$ near its zero: $f(z)=(z-z_j)/\chi_j+O((z-z_j)^2)$. The  magnetization  rotates (counter) clockwise around the center of the vortex if its $Im~ \chi_j>0$ (resp. $<0$).

In terms of these definitions, the following constraints must always hold for any doubly-periodic function $f(z)$:
\begin{enumerate}[label=\textnormal{\arabic*)}]
 \item \label{c:zero-charge} The number of vortices equals to the number of anti-vortices (saddles) in each
elementary cell.
 \item \label{c:positions} The sum of positions of vortices  in a cell equals to the sum of positions of
saddles modulo $L$.
 \item \label{c:chirality} The sum of chiralities $\chi_j$ of all vortices inside the elementary cell equals zero.
\end{enumerate}

\begin{figure}
\begin{picture}(70, 30)
\linethickness{.5mm}
\thinlines
\put(5,5){\vector(1,0){50}}
\put(5,5){\vector(1,3){10}}
\put(15,35){\line(1,0){50}}
\put(55,5){\line(1,3){10}}

\put(12,34){$\tau$}
\put(57,4){$1$}

\put(25,5){\circle*{1}}
\put(10,20){\circle*{1}}
\put(35,35){\circle*{1}}
\put(60,20){\circle*{1}}

\put(45,15){\circle*{1}}
\put(30,20){\circle*{1}}

\put(20,13){\circle{1}}
\put(35,13){\circle{1}}
\put(25,25){\circle{1}}
\put(50,25){\circle{1}}

\put(29,22){$p_j$}
\put(34,10){$z_j$}

\thicklines 
\multiput(10,0)(5,0){10}{\line(1,0){3}}
\multiput(10,0)(2.5,7.5){4}{\line(1,3){1.2}}
\multiput(20,30)(5,0){10}{\line(1,0){3}}
\multiput(60,0)(2.5,7.5){4}{\line(1,3){1.2}}

\put(35,0){\vector(1,0){3}}
\put(48,30){\vector(-1,0){3}}

\end{picture}
\caption{\label{fig:lattice}Integration path (dashed parallelogram) around elementary cell of the 2-d lattice, formed by the complex vectors $1$ and $\tau$, avoiding zeros $z_j$ and poles $p_j$ with $j=1..n$ of a degree $n=4$ function.}
\end{figure}
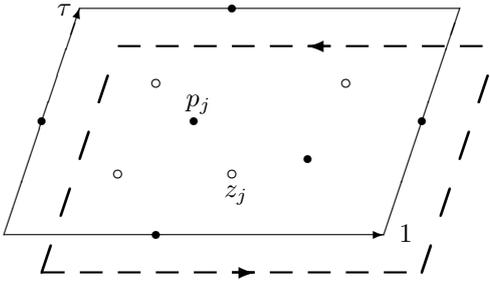

{\bf Proof.} These constraints directly follow from basic properties of elliptic functions
\cite{akhiezer1990translation}.
Let us slightly shift the elementary parallelogram $\Pi$ so that $f(u)$ has neither
poles nor zeros  on its boundary like that shown in Fig~\ref{fig:lattice}. Then
the constraint \ref{c:zero-charge} follows from
\begin{equation}
 \label{eq:p1}
 0=\oint_{\partial\Pi} \ud\, \log f(z)=2\pi \imath (\# [f(z)] -  \# [1/f(z)]),
\end{equation}
where $\# [f(z)]$ denotes the number of zeros (including their multiplicities) of the function $f(z)$ inside $\Pi$ and $\# [1/f(z)]$, accordingly, the number of poles. Indeed, on the opposite sides of $\Pi$ we integrate the same form in opposite directions, this gives the equality on the left, the right equality is due to the residue formula. The constraint \ref{c:positions} follows from
\begin{equation}
 \label{eq:p2}
2\pi \imath~L\ni\oint_{\partial\Pi} \ud (z\log f(z))=
2\pi \imath (\sum_{j=1}^n z_j -  \sum_{j=1}^n p_j),
\end{equation}
where $z_j$ ( $p_j$ ) are the positions of zeros (poles) of function $f(z)$ inside $\Pi$ repeated if the zero (pole) has a multiplicity. The sum of integrals over the vertical (inclined) sides of $\Pi$ in Fig~\ref{fig:lattice} equals to $\tau$ times increment of $\log f(z)$ on such a side. The sum of integrals over the horizontal sides belongs to $2\pi \imath\mathbb{Z}$ for the same reason. Furthermore, for the same reason as (\ref{eq:p1}),
\begin{equation}
 \label{eq:p3}
 0=\oint_{\partial\Pi} \ud z/f(z)=2\pi \imath \sum_{j=1}^n \chi_j , 
\end{equation}
proving the constraint \ref{c:chirality}. It follows from \ref{c:chirality} that the magnetization rotation around all vortices cannot be clockwise. Also, all vortices cannot be only sinks (attractors) or only sources (repellors), there should be vortices of both types or neutral pure vortices with $Re~ \chi_j=0$.

Constraints \ref{c:zero-charge} and \ref{c:positions} above are necessary and sufficient for the existence of a doubly periodic function with the given set of zeros and poles. This function is unique up to multiplication by a nonzero constant and may be represented in terms of elliptic theta (or its close relative sigma) function \cite{akhiezer1990translation}.
The elliptic function $\theta_1(z|\tau)$ is an odd function of variable $z$ that has zeros exactly in the points of the period lattice $L$. It acquires simple factors when the lattice vectors are added to the argument: $\theta_1(z+1|\tau) =  -\theta_1(z|\tau)$, $\theta_1(z+\tau|\tau) = -\exp(-\imath\pi\tau- 2\pi \imath z)\theta_1(z|\tau)$.

Now one can easily check that if the set of zeros $\{z_j\}_{j=1}^n$ and poles $\{p_j\}_{j=1}^n$ (points may collide in each set) satisfy condition 2), namely $\sum_{j=1}^n z_j-\sum_{j=1}^n p_j=M+N\tau ~~\in L(\tau)$, then the following combination
\begin{equation}
f(z) = \exp(-2\pi \imath N z)\prod_{j=1}^n\frac{\theta_1(z-z_j)}{\theta_1(z-p_j)}
\label{repre} 
\end{equation}
is periodic with respect to the lattice and has prescribed sets of zeros and poles
modulo the lattice. Moreover, this function is essentially unique, for otherwise we
consider the ratio of two such functions. This ratio will be a doubly periodic
function without zeros and poles, hence a constant by the maximum principle.

\begin{figure}[thb]
\centering
\includegraphics[width=\columnwidth]{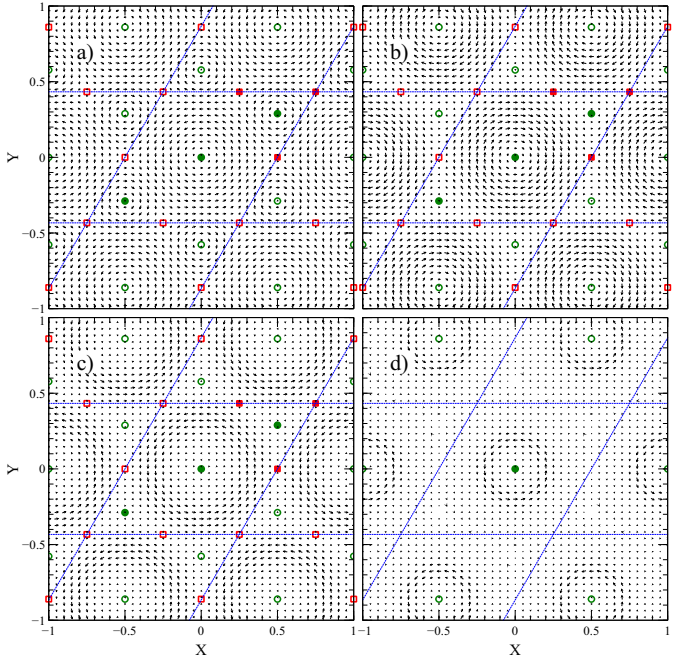}
\caption{\label{fig:hexagonal} In-plane components of the magnetization vectors of a thin film with doubly-periodic magnetization textures in a hexagonal lattice with lattice vectors $1$ and $\tau=\exp (\imath\pi/3)$. Thin lines separate the lattice cells. Subfigure a) plots the Eq. (\ref{repre})  with $n=3$, vortices at $z_j = 0, 1/3+\tau/3, 2/3+2\tau/3$ and saddles at $p_j = 1/2,\tau/2, 1/2+\tau/2$, shown by the filled circles and squares (the open ones are their translations by the integer multiples of the lattice vectors). Subfigures b) and c) show the linear superposition of spirals and densely packed magnetic bubble lattice, as described in the text. Subfigure d) shows sparse magnetic bubble lattice. The actual numerical positions of vortices and saddles in subfigures a)-c) match exactly. There are no noticeable saddles in d), the magnetization is largely perpendicular to the film (figure) plane outside of the domain walls.}
\end{figure}

Let us now attempt to faithfully reproduce the real-space image of the hexagonal skyrmion lattice from the Ref.~\onlinecite{yu2010realspaceobservation}. From the property \ref{c:zero-charge} it follows that a hexagonal lattice can't consist of only vortices. Thus, to respect the hexagonal symmetry we need to add at the very least six antivortices around the vortex at the corners of the unit cell (each of them belongs to several unit cells with three full antivortices per unit cell). To balance these antivortices (as per constraint \ref{c:zero-charge}) two more full vortices in a unit cell are necessary. Placing these vortices and antivortices in accordance with the constraint \ref{c:positions} results in the magnetization texture shown in Fig.~\ref{fig:hexagonal}a.
Please note that the relative size of the vortices as well as their relative chirality was not specified when constructing the image, it follows emergently from the  Eq.~\ref{repre} and the prescribed positions of zeros and poles.

A similar structure can be imagined as a superposition of three spirals \cite{MuhlbauerSkyrmionLattice} shown in Fig.~\ref{fig:hexagonal}b, which is, probably, the closest to the experimental observations \cite{yu2010realspaceobservation}. Specifically, the in-plane components of the magnetization vectors are given by the complex function $w_\mathrm{S}(z,\zc)=\imath q_1 \cos (q \uRe \overline{q_1} z ) + \imath q_2 \cos (q \uRe \overline{q_2} z ) + \imath q_3 \cos (q \uRe \overline{q_3} z )$, where $q_1=\exp (\imath \pi/2), q_2=exp(-\imath 5 \pi /6), q_3=exp(-\imath \pi /6)$ are the wave vectors of the three spirals and $q=2\pi\cdot2/\sqrt{3}$ with the specific angles in $q_i$ and factor in $q$ selected to match the lattice used in the previous example.

Finally, the skyrmion lattice can be represented as a dense hexagonal lattice of magnetic bubble domains, 
separated by Bloch domain walls. We can model the in-plane magnetization components profile of a single Bloch domain wall by a complex function $w_\mathrm{B}(z,\zc)=\imath z/|z| \sech (2 (|z|-s)/d)$ with $s$ and $d$ specifying the domain size and the domain wall width. This function is exponentially sharp as it is typical for the 180\degree{} Bloch domain walls in uniaxial thin film ferromagnets. Arranging these walls into a hexagonal lattice, which is identical to the previous example, we get the magnetization texture, shown in Fig.~\ref{fig:hexagonal}c.

As one can verify directly, the positions of zeros of the in-plane components of the magnetization for all three considered models coincide exactly and satisfy the constraints we have formulated above. It is worth noting that the models are very different. The constraints follow from considering the exchange energy term only, but are also satisfied by a densely packed bubble lattice with Bloch walls (including the case of non-negligible uniaxial magnetic anisotropy) and the model for an A-phase in MnSi, which specifically takes into account the Dzyaloshinskii-Moria interaction. This shows the defining role of the magnetization vector field continuity, mediated by the exchange interaction, which ultimately forces the magnetization to reflect the topology of space.

As we can see in Fig.~\ref{fig:hexagonal}, different magnetic interactions and the external factors may distort the magnetization texture (and, in particular, introduce quantitative differences between the skyrmions and bubbles \cite{ISZ90}), but not necessarily destroy its topological features. However, this is not always the case. In the case of low-density magnetic bubble domain lattice, shown in Fig.~\ref{fig:hexagonal}d, the constraints \ref{c:zero-charge}-\ref{c:chirality} do not apply. The magnetization far enough outside of the bubbles has no appreciable in-plane components, the positions of saddles and vortices (besides the ones, corresponding to the bubble centers) are not defined. This is because the in-plane phase (azimuthal angle) of the magnetization vector far enough from the bubble loses its meaning. There is no necessity to match this phase continuously and globally across the lattice (which is the physical reason behind the existence of the topological constraints) and thus the individual bubbles in a low-density bubble lattice gradually (as its density decreases) become unconstrained globally, making it possible to move (create, annihilate) them individually, as well as individually control their chiralities (see e.g. Figs.~3b and 3c in Ref.~\onlinecite{Nagaosa2013} for examples of such magnetic bubble lattices).

A continuous transition from a more constrained dynamics to the less constrained as scale changes is not new to physics. It reminds the transition from quantum to classical mechanics. In the former case the phase of the complex wave-function is important, which implies a number of non-obvious physical effects, in the latter case only the amplitude (whether the particle is present or not) matters. On this language, the low-density magnetic bubble lattice is a classical object, which was already used in the last century as a basis for classical computations \cite{eschenfelder1980}. Topological solitons, on the contrary, are similar to quantum objects, endowed with their own version of ``spooky action at a distance'' in the form of the topological constraints (following, as in the case here, from purely local interactions of classical magnetic moments). Consider e.g. a constraint \ref{c:positions}, which means that moving the vortex/anti-vortex in one location, implies that something must move in the other distant locations as well (in practice the distortion will, of course, propagate locally across the film until all the texture matches in a new static configuration, satisfying the constraints). This hints at a possibility to create a version of quantum computer using topological solitons for computation.

In summary, we have derived a set of topological constraints for the periodic magnetization textures in thin ferromagnetic films. The constraints are the result of the necessity to match the magnetization vector in-plane phase across the whole film. We propose to use these constraints to distinguish between the skyrmion magnetization textures (where the constraints are relevant) and the magnetic domains (magnetic bubbles specifically, stabilized by the easy-axis anisotropy), where the in-plane phase information is lost inside the uniformly magnetized domains (and the topological constraints become irrelevant). Skyrmion textures promise new topological phenomena due to their global geometric phase (Berry's phase) effects. The magnetic domains (such as magnetic bubbles), on the other hand, are easier to manipulate individually. The easy axis anisotropy, favouring the uniform domains, is a factor, suppressing the geometric phase. Finite magnetically soft (when the easy-plane magnetostatic anisotropy dominates) planar nanoelements can support even more sophisticated topological constraints \cite{BM17}.

The support of the Russian Foundation of Basic Research under the project {\tt RFBR 16-01-00568} is acknowledged.

\bibliographystyle{elsarticle-num}

\end{document}